\documentclass[12pt]{article}
\usepackage{a4,amsmath,lscape,epsfig}
\def\ii{\'{\i}}
\def\beq{\begin{equation}}
\def\eeq{\end{equation}}
\def\beqa{\begin{eqnarray}}
\def\eeqa{\end{eqnarray}}
\def\ban{\begin{eqnarray*}}
\def\ean{\end{eqnarray*}}
\def\bi{\begin{itemize}}
\def\ei{\end{itemize}}

\def\d{\mbox{d}}

\begin{document}

\begin{center}
{\bf Relativistic Thomas-Fermi description of Sm isotopes at finite 
temperature}\\
\end{center}

\begin{center}
{\it C. Provid\^encia $^1$, D.P.Menezes $^2$ and L. Brito $^1$}\\
{\it $^1$ Centro de F\ii sica Te\'orica - Dep. de F\ii sica -
  Universidade de Coimbra}\\
{\it P-3004 - 516 Coimbra - Portugal}\\
{\it $^2$  Dep. de F\'{\i}sica - CFM -
  Universidade Federal de Santa Catarina}\\
{\it Florian\'opolis - SC - CP. 476 - CEP 88.040 - 900 - Brazil}\\
\end{center}

\vspace{0.50cm}

\begin{abstract}
{The caloric curve (excitation energy per particle as a function
of temperature) for finite nuclei is calculated within the non--linear
Walecka model for different proton fractions and different 
parameterizations. The results obtained are compared with 
published experimental data and other theoretical results. 
Other properties related with
the droplet formation as the surface energy, neutron skin thickness
and binding energy per nucleon are also displayed as a function 
of temperature
and their behaviours are discussed. It is shown that the
caloric curve is sensitive to the proton fraction and to the Coulomb 
interaction. At $T=0$ the droplet properties do not depend on the 
parameterization used. At finite temperature however, the sensitivity 
appears.}
\end{abstract}

\vspace{0.50cm}
PACS number(s): {\bf 21.10.-k, 21.30.-x,21.65.+f,25.70.-z}
\vspace{0.50cm}

\newpage

\section{Introduction}

One of the most important problems in contemporary nuclear physics and 
astrophysics is the determination of the properties of nuclear matter as 
functions of density, temperature and the neutron-proton composition. 
This information is important in understanding the explosion mechanism of
supernova, the cooling rate of neutron stars or the processes involved 
in the formation of the trans-iron elements
\cite{pr,bck,kahana,lat91,sum94,cosmos}.

Recent advances in experiments using heavy-ions at high energies and  
radioactive beams with 
large neutron or proton excess have made it possible to create
not only nuclei at the limits of stability, but also a transient state 
of nuclear matter with appreciable isospin asymmetry, thermal excitation, 
and compression \cite{exp1,exp2,yen96}.

The production of several intermediate mass fragments in a short time scale
during heavy ion collisions is known as nuclear multifragmentation. 
In multifragmentation experiments, an equilibrated system is always formed
and the behaviour of the energy fluctuations suggests that the system
undergoes a liquid--gas phase transition. One of the evidences of this
transition is the fact that the heat--capacity exhibits a peak at a certain
temperature. Although 
different pictures are normally used in order to describe this phase 
transition phenomenon \cite{Gross,INDRA,lbct}, some common features are 
present in all of them. 
The caloric equation of state, which is given by the 
excitation energy per nucleon in terms of the thermodynamic temperature is
an important quantity to be investigated in the search for a phase transition.
Notice that the temperature at which phase transitions take place 
experimentally is not the same as the thermodynamic one.  

In order to explain recent experimental results, 
one must study not only the ground and 
excited states of normal nuclei, but also nuclear states of high excitation 
and far from stability. In particular it is important to understand 
the role of the isospin degree of freedom
in heavy-ion collisions at intermediate energies and  
 the properties of nuclear matter in the 
region between symmetric nuclear and pure neutron matters. 

The radius, thickness of neutron skins, deformation, 
binding energy, density distributions, and other properties 
of radioactive nuclei near the
drip lines depend sensitively on the isospin-dependence of the
nuclear equation of state ({\sc eos}). The authors of references 
\cite{tanihata} and \cite{LKW97} have recently stressed the
possibility of extracting the  {\sc eos} of asymmetric nuclear matter through
the investigation of these properties as well as via reactions of 
neutron-rich nuclei at intermediate energies. 

The properties of neutron rich nuclei have been studied using, among others,
Skyrme-Hartree-Fock ({\sc shf}) and the 
relativistic mean-field  ({\sc rmf})  theories \cite{SW}.  
In particular it has been 
shown by Tanihata \cite{tanihata} that in {\sc shf} the saturation density 
essentially does not change, 
but in {\sc rmf} it decreases rapidly as nuclear matter becomes more 
neutron-rich. Most {\sc eos} used in the description of neutron stars
and supernova are obtained from nonrelativistic models. It is
therefore of interest to study the {\sc eos}, namely the inhomogeneous 
phase at the liquid-gas phase
transition, using relativistic models. The liquid-gas phase transition 
is probed not only by investigating the properties of neutron stars but 
also through experimental results obtained in collisions involving  
heavy-ions and in
multifragmentation experiments, already mentioned. 

Nucleation process, which is known as the liquid droplet formation 
in a gas background, is a very old problem with many applications in
different areas of physics and chemistry. Newton (1687) and Laplace (1816)
dedicated themselves to understand the physics underlying the nucleation
theory earlier than the formulation of the zeroth law of thermodynamics
in 1913 \cite{history}. In this work we turn into the nucleation process 
inside  nuclear matter.

Within the framework of relativistic models, the liquid-gas phase transition 
in nuclear matter has been investigated at zero and finite temperatures for 
symmetric and asymmetric semi-infinite systems \cite{rs,md,epsw,cev,toki98}.
Droplet formation in  the liquid-gas phase 
transition in cold~\cite{mp1,nosso3} and hot~\cite{mp2} asymmetric nuclear 
matter 
in the context of relativistic mean field theory, namely using  the 
non-linear Walecka model
(NLWM)~\cite{bb} has been studied recently using a Thomas Fermi approximation.
As shown in Refs.~\cite{mp1,mp2}, the optimal 
nuclear size of a droplet in a neutron gas is determined by a delicate balance
between nuclear Coulomb and surface energies. The surface energy favors nuclei
with a large number of nucleons $A$, while the nuclear Coulomb self-energy 
favors small nuclei. 

In a previous work \cite{debcp}, the influence 
of the proton fraction in the caloric curve was calculated for nuclei 
obtained within the framework of the approach mentioned above. 
In studying multifragmentation,
an input parameter called the freeze-out volume, which simulates a phase 
transition at constant volume, is normally used 
\cite{Gross}. The consequences for the caloric curve of imposing 
thermalization in a freeze-out volume have also been discussed.

The present work extends our study, by investigating the influence 
of the temperature in some important quantities related 
to the droplet formation, such as the surface energy, the neutron 
skin thickness and the binding energy per nucleon.
The importance of the Coulomb interaction and the nucleus
proton fraction is investigated. 
The effects of using different parameterizations are also analysed. 
In this way, a more complete understanding 
of the caloric curve is 
obtained.

The paper is organized as follows.
In section 2 the thermodynamical potential in the framework of 
the Thomas-Fermi
approximation is calculated for the NLWM and the two--phase coexistence
is discussed. In section 3 we present the numerical 
results. Finally, in the last section the conclusions are drawn.

\section {The Thomas-Fermi Approximation in the Extended Non Linear Walecka 
Model}

In what follows we describe the equation of state of asymmetric matter 
within the framework of the
relativistic non-linear Walecka model \cite{bb}, \cite{ring}
with the inclusion of 
$\rho$-mesons and the electromagnetic field. 
In this  model the nucleons are coupled to scalar-isoscalar $\phi$, 
vector-isoscalar $V^\mu$, vector-isovector $\vec b^\mu$  meson fields and the
electromagnetic field $A^\mu$.
The lagrangian density reads:
$$
{\cal L}=\bar \psi\left[\gamma_\mu\left(i\partial^{\mu}-g_v V^{\mu}-
\frac{g_{\rho}}{2}  \vec{\tau} \cdot \vec{b}^\mu 
-e A^\mu \frac {(1+ \tau_3)}{2}\right) 
-(M-g_s \phi)\right]\psi
$$
$$
+\frac{1}{2}(\partial_{\mu}\phi\partial^{\mu}\phi
-m_s^2 \phi^2) - \frac{1}{3!}\kappa \phi^3 -\frac{1}{4!}\lambda
\phi^4
-\frac{1}{4}\Omega_{\mu\nu}\Omega^{\mu\nu}+\frac{1}{2}
m_v^2 V_{\mu}V^{\mu}
$$
\begin{equation}
+\frac{1}{4!}\xi g_v^4 (V_{\mu}V^{\mu})^2 -\frac{1}{4}
\vec B_{\mu\nu}\cdot\vec 
B^{\mu\nu}+\frac{1}{2}
m_\rho^2 \vec b_{\mu}\cdot \vec b^{\mu}  
-\frac{1}{4}F_{\mu\nu}F^{\mu\nu}\;,
\end{equation}
where
$\Omega_{\mu\nu}=\partial_{\mu}V_{\nu}-\partial_{\nu}V_{\mu}$ ,
$\vec B_{\mu\nu}=\partial_{\mu}\vec b_{\nu}-\partial_{\nu} \vec b_{\mu}
- g_\rho (\vec b_\mu \times \vec b_\nu)$ 
and
$F_{\mu\nu}=\partial_{\mu}A_{\nu}-\partial_{\nu}A_{\mu}$.
The model comprises the following parameters:
three coupling constants $g_s$, $g_v$ and $g_{\rho}$ of the mesons to
the nucleons, the nucleon mass $M$, the masses of
the mesons $m_s$, $m_v$, $m_{\rho}$, the
electromagnetic coupling constant $e=\sqrt{\frac{4 \pi}{137}}$ and
the self-interacting coupling constants $\kappa$, $\lambda$ and $\xi$. 
In this work, unless otherwise stated, we use a set of constants
usually identified as NL1  
\cite{sed},
which is not unique, but it gives a good description 
of the ground state properties
of many stable nuclei. The consequences of using different 
parameterizations
is also investigated and, for this purpose, 
we use two other sets known as NL3 \cite{nl3} and TM1 \cite{tm1}.
The values of the constants are displayed in table 1 and the most 
important bulk properties obtained with these three parameterizations are
displayed in table 1a. 

The basic quantity in the Thomas--Fermi approximation is the phase--space
distribution function for protons and neutrons:
\begin{equation}
f_{i\pm}({\mathbf r},{\mathbf p},t)\,=\,
\frac{1}{1+\exp[(\epsilon\mp\nu_i)/T]}\;
, \quad i=p,n,\end{equation}
where  $\nu_i\;=\mu_i-{\cal V}_{i0}$ are the effective chemical potentials
with $\mu_i$ being the chemical potentials for particles of type $i$,
$$
{\cal V}_{p0}= g_v V_0  + \frac{g_\rho}{2} b_0 + e A_0\; ,
\quad {\cal V}_{n0}= g_v V_0  - \frac{g_\rho}{2}  b_0 \; ,
$$
$\epsilon=\sqrt{p^2+{M^*}^2}$,
$M^* =M-g_s\phi$ is the effective nucleon mass and $T$ is the temperature. 

The classical entropy of a Fermi gas is given by 
\begin{equation}
S=-2\sum_{i=p,n}\int\frac{\d^3r\d^3p}{(2\pi)^3}\,\left(f_{i+} 
\ln\left(\frac{f_{i+}}{1-f_{i+}}\right)+
\ln(1-f_{i+}) +(f_{i+}\leftrightarrow f_{i-})\right)\;,
\end{equation}
and the thermodynamic potential is defined as

\begin{equation}
\Omega\,=\,E-TS-\sum_{i=p,n}\mu_i N_i,\; 
\label{Omega}\end{equation}
where $N_p\,, \, N_n$ are, respectively, the proton and the neutron number:
\begin{equation}
N_i\,=\,\int\d^3r \rho_i({\mathbf r},t),\qquad \rho_i=2
\int\frac{\d^3p}{(2\pi)^3}(f_{i+}-f_{i-}), \quad i=p,n\; .
\label{rhoi}\end{equation}
\noindent From the above expressions 
we get for (\ref{Omega}), in the static approximation,
\beq
\Omega=  \int\d^3r \left(\frac{1}{2}\left [
(\nabla \phi)^2 -(\nabla V_0)^2 - (\nabla b_0)^2 - (\nabla A_0)^2
\right] - V_{ef}\right)
\end{equation}
with
$$
V_{ef}=
-\frac{1}{2} \left[ 
m_s^2 \phi^2 + \frac{2}{3!} \kappa \phi^3 + \frac{2}{4!} \lambda \phi^4
-m_v^2 V_0^2 -\frac{2}{4!} \xi g_v^4 V_0^4 -m_\rho^2 b_0^2  \right]$$
\begin{equation}
+ 2 T \sum_i \int \frac{\d^3p}{(2\pi)^3} \left[  
\ln (1 + e^{-(\epsilon - \nu_i)/T})
+ \ln (1 + e^{-(\epsilon + \nu_i)/T}) \right] \label{vef}.
\end{equation}
The energy density reads
$${\cal E}(r,T)=\,2 \sum_i \int\frac{d^3p}{(2\pi)^3}\,
\left[\epsilon(f_{i+}+f_{i-}) +{\cal V}_{i0}(f_{i+}-f_{i-})
\right]$$
$$+\frac{1}{2}\left [
(\nabla \phi)^2 -(\nabla V_0)^2 - (\nabla b_0)^2 - (\nabla A_0)^2
\right] +$$
\beq
+\frac{1}{2} \left[ 
m_s^2 \phi^2 + \frac{2}{3!} \kappa \phi^3 + \frac{2}{4!} \lambda \phi^4
-m_v^2 V_0^2 -\frac{2}{4!} \xi g_v^4 V_0^4  
-m_\rho^2 b_0^2  \right]. \label{dens}
\end{equation}
For later discussion it is convenient to separate ${\cal E}(r,T)$ into  two
parts: the nuclear contribution, ${\cal E}_N$, and the 
electromagnetic one; the latter is
$${\cal E}_{\rm coul.}=\,2  \int\frac{d^3p}{(2\pi)^3}\,
e\,A_0\,(f_{p+}-f_{p-})-\frac{1}{2}(\nabla A_0)^2\,.
$$

The fields that minimize $\Omega$ satisfy the equations
\begin{equation}
\frac{\partial V_{ef}}{\partial\phi}=-m_s^2\phi
-\frac{1}{2}\kappa \phi^2 -\frac{1}{3!} \lambda\phi^3 + g_s \rho_s
=-\nabla^2 \phi
,\label{phi} \end{equation}
\begin{equation}\frac{\partial V_{ef}}{\partial V_0} =m_v^2 V_0 
 +\frac{1}{3!} \xi g_v^4 V_0^3 - g_v \rho_B =\nabla^2 V_0
, \label{V0}\end{equation}
\begin{equation}\frac{\partial V_{ef}}{\partial b_0} =m_\rho^2 b_0
-\frac{g_\rho}{2} \rho_3=\nabla^2 b_0
, \label{b0}\end{equation}
\begin{equation}\frac{\partial V_{ef}}{\partial A_0}=-e \rho_p
=\nabla^2 A_0
,\label{A0}\end{equation}
where $\rho_B=\rho_p+\rho_n,\, \rho_3=\rho_p-\rho_n$ 
and 
$$\rho_s= 2 \sum_{i=p,n}
\int \frac{\d^3p}{(2\pi)^3}
\frac{M^*}{\epsilon}\left(f_{i+}+f_{i-}\right).$$

These coupled differential equations are solved  numerically 
and all relevant quantities which depend on the fields are calculated. 
Looking for phase transitions in binary systems such as this one requires
the study of three kinds of instabilities which can occur: 
mechanical, diffusive and thermodynamical.  
The condition for mechanical stability requires
\beq
\left(\frac{\partial P}{ \partial\rho_B}\right)_{Y_p, T}\ge 0\;,
\end{equation}
where $P$ is the pressure and $Y_p=\rho_p/\rho_B$ is the proton fraction. 
The condition for diffusive stability implies the 
inequalities
\beq
\left( \frac{\partial \mu_p}{\partial Y_p} \right)_{P,T} \ge 0
~~~{\rm and} ~~~
\left( \frac{\partial \mu_n}{\partial Y_p} \right)_{ P,T} \le 0\;.
\label{ds}
\end{equation}
These conditions reflect the fact that in a stable system, energy is required 
to increase
the proton concentration 
at constant pressure and temperature. 
Thermodynamical stability is expressed by 
\beq
c_v=\left(\frac {d\varepsilon^*}{dT}\right)_{v, Y_p}\ge 0, \label{sh}
\end{equation}
where $c_v$ is the specific heat and
\beq
\varepsilon^*=\varepsilon(T)-\varepsilon(T=0), \label{excen}
\end{equation}
is the excitation energy per particle. The
total energy per particle at temperature $T$ is given by \cite{shlomo1}
\beq
\varepsilon(T)=\int \frac{{\cal E}(r,T)}{A} d^3r\,=\, 
\varepsilon_{N}(T) + \varepsilon_{\rm coul.}(T)\,,\label{et}
\end{equation}
where $A$ is the total number of particles in the volume under consideration,
defined at the end of section 3, and $\varepsilon_{N}(T)$,  
$\varepsilon_{\rm coul.}(T)$ denote, respectively, the nuclear and 
electromagnetic contributions.

The two-phase liquid-gas coexistence is governed by the Gibbs condition
\beqa
\mu_i(\rho_p,\rho_n,M^*)&=&\mu_i(\rho_p^{\prime},\rho_n^{\prime},
{M^*}^{\prime}),~~i=p,n
\label{gib1}\\
{ P}(\rho_p,\rho_n,M^*)&=&
{ P}(\rho_p',\rho_n',{M^*}^{\prime})\; ,
\label{gib2}
\eeqa
where the primed and unprimed quantities correspond to the two different 
phases.

In the mean field approximation for infinite nuclear matter, 
the meson fields are replaced by their expectation values. From the equations
of motion, they can easily be obtained \cite{mp1}
and the thermodynamic quantities of interest are given in terms of these
meson fields. 
We have made use of the geometrical construction~\cite{barranco,ms} in order 
to obtain the chemical potentials in the two coexisting phases for each 
pressure of interest.
In order to obtain the binodal section which contains points under
the same pressure for different proton fractions, we have solved
eqs. (\ref{gib1}) and (\ref{gib2}) simultaneously 
with the following ones:
\begin{equation}
m_s^2 \phi_0 + \frac{\kappa}{2} \phi_0^2 
+\frac{\lambda}{6}\phi_0^3 = g_s \rho_s(\nu_p,\nu_n,M^*)
\end{equation}
and
\begin{equation}
m_s^2{\phi_0}^{\prime} + \frac{\kappa}{2} {{\phi_0}
^{\prime}}^2 
+\frac{\lambda}{6}{{\phi_0}^{\prime}}^3 = g_s \rho_s(\nu_p^{\prime},
\nu_n^{\prime},{M^*}^{\prime}).
\end{equation}

\section{Numerical Results for Finite Systems}

The solution for the infinite system gives us the initial and
boundary conditions for the program which integrates the set of
coupled non-linear differential equations (\ref{phi}) to (\ref{A0})
in the Thomas-Fermi approximation. 
In this work the numerical
calculation was carried out with the iteration procedure described in
Refs.~\cite{mp1,nosso3,mp2}.

Some quantities of interest are outlined below.
The surface energy per unit area of the droplets in the small
thickness approximation is
\begin{equation}
\sigma=\int_0^\infty \d r \left[ \left(\frac{\d \phi}{\d r}\right)^2- 
\left(\frac{\d V_0}{\d r}\right)^2 -
\left(\frac{\d b_0}{\d r}\right)^2 
\right].
\label{sig}\end{equation}

The proton and neutron radii in 
the spherical geometry, $R_i~(i=p,n)$, are
defined as 
\begin{equation}
\int_0^{R'} \rho_i(r) r^2 d r = \frac{1}{3}
\left[\rho_{i,l}R_i^3 + \rho_{i,g}({R'}^3-R_i^3) \right],
\label{raiosp}
\end{equation}
where $\rho_{i,l}$ and $\rho_{i,g}$  refer to the liquid and gas density
respectively and $R'$ is the value 
of $r$ for which the fields and density reach their asymptotic gas values.
 
The {\it neutron skin thickness} 
is defined as~\cite{cev} 
\begin{equation}
\Theta=R_n-R_p.
\end{equation}
Other quantities of interest are the number of protons ($Z$) and neutrons 
($N$) in the droplet, given by
\begin{equation}
Z=4 \pi\int_0^{R'} \rho_p(r) r^2 d r - \frac{4\pi}{3}{R'}^3\rho_{p,g}
\end{equation}
and 
\begin{equation}
N=4 \pi\int_0^{R'} \rho_n(r) r^2 d r - \frac{4\pi}{3}{R'}^3\rho_{n,g} .
\end{equation}

The binding energy per nucleon is
\beq
\frac{B}{A}=\varepsilon(T)-M, 
\end{equation}
where $A= Z+N$.

In the sequel we 
mainly study two nuclei: $^{150}_{62}{\rm 
Sm}_{88}$ and $^{166}_{62}{\rm Sm}_{104}$. We have chosen these two 
nuclei because they
lie in the mass range of interest for intermediate energy
heavy-ion collisions
\cite{exp1,exp2,hagel}. On the other hand, two 
isotopes with quite different number of neutrons
are taken in
order to study the effect of proton--neutron asymmetry.
 
\begin{figure}
\begin{center}
\epsfig{file=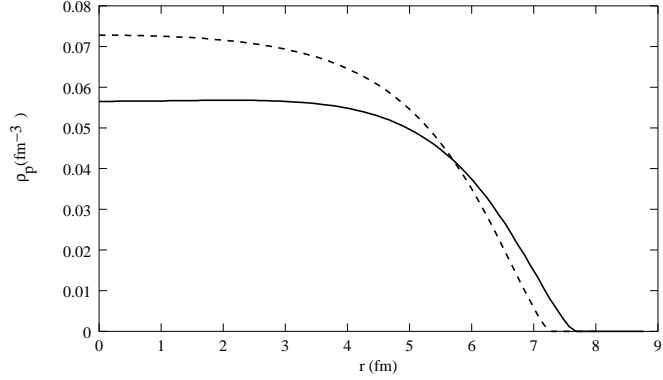,height=10cm,angle=270}
\caption{The  $^{150}_{62}{\rm Sm}_{88}$ ($Y_p=0.41$)
density profiles for protons $\rho_p(r)$  in fm$^{-3}$ are plotted
with (solid) and without (dashed) the inclusion of the 
electromagnetic field for $T=0$ MeV.}  
\label{fig1}
\end{center}
\end{figure}

In figure \ref{fig1}, the proton density profiles are plotted 
with and without the 
electromagnetic field for the  $^{150}_{62}{\rm Sm}_{88}$ 
at $T=0$. The central
density is smaller when the Coulomb interaction is 
considered because more
protons are pushed off to the surface. The same behaviour is obtained for
any other fixed temperature, whenever both systems, with and without the 
inclusion of the electromagnetic field are found. Notice, however, 
that the critical temperature, from which the droplet ceases to exist,
is smaller when the electromagnetic field is included, as compared with
systems where the Coulomb interaction is disregarded. 

\begin{figure}
\begin{center}
\epsfig{file=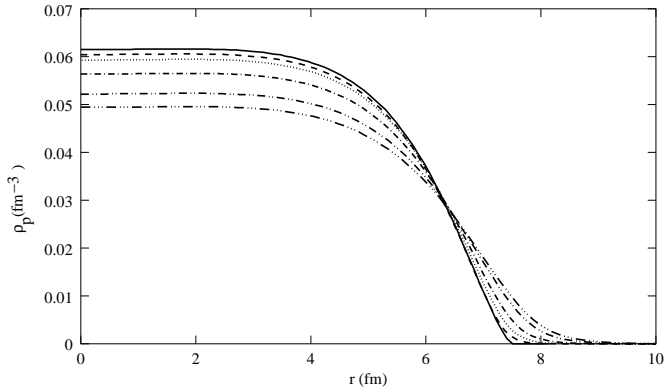,height=10cm,angle=270}
\caption{The  $^{150}_{62}{\rm Sm}_{88}$ ($Y_p=0.41$)
density profiles for the protons $\rho_p(r)$  in fm$^{-3}$ are plotted
with the inclusion of the electromagnetic field
for temperatures varying from $T=0$ 
(at the top, on the left) to $T=6.5$ MeV (at the 
bottom, on the left).}  
\label{fig2}
\end{center}
\end{figure}

In figure \ref{fig2}, the $^{150}_{62}{\rm Sm}_{88}$ density profiles 
for the protons are 
plotted with the inclusion of the electromagnetic field,
for temperatures varying from $T=0$ to $T=6.5$ MeV. 
The central density decreases 
with the increase of the temperature and consequently, 
the hotter the droplet,
the larger it is. The same behaviour is observed 
when the Coulomb interaction
is not included and for any other proton fraction. 
These results are in accordance with those 
obtained in \cite{shlomo2}, 
with a non--relativistic model.
 
\begin{figure}
\begin{center}
\epsfig{file=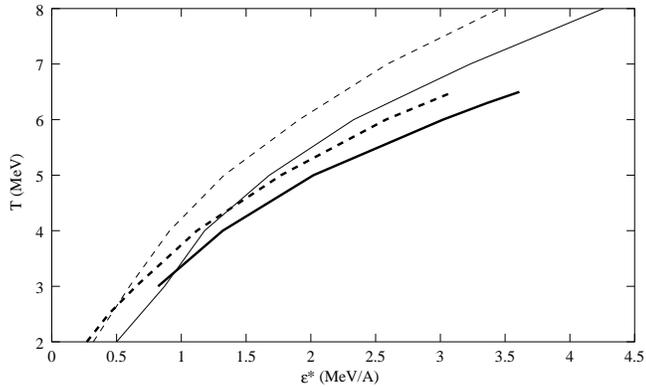,height=10cm,angle=270}
\caption{The caloric curves are shown for 
$^{150}_{62}{\rm Sm}_{88}$ ($Y_p=0.41$) with (thick solid line) and
without (thin solid line) the Coulomb 
interaction and for $^{166}_{62}{\rm Sm}_{104}$ 
($Y_p=0.37$) with (thick dashed line) and without (thin dashed 
line) the Coulomb interaction.}
\label{fig3}
\end{center}
\end{figure}

In table 2, we show the surface energy, the proton radius, the neutron
skin thickness, the central density, the binding energy per nucleon
and the excitation energy per particle calculated without the inclusion 
of the Coulomb interaction for the $^{150}_{62}{\rm Sm}_{88}$, 
which has a proton
fraction equal to $0.41$. In table 3, the same quantities are
shown when calculated with the electromagnetic interaction included.
In tables 4 and 5, these
quantities are obtained for the  $^{166}_{62}{\rm Sm}_{104}$,
which has a proton
fraction of $0.37$, without and with the Coulomb interaction
respectively.
Examining these tables, several common trends 
are noticed, despite the difference in the proton fraction. In general, 
the surface energy, the central density and the binding 
energy decrease with the increase of the temperature. 
Inversely, the proton radius and the excitation energy per particle 
increase as the temperature becomes higher. 
Comparing tables 2 with 3 
and 4 with 5, one observes that the non inclusion of the Coulomb 
interaction
gives rise to smaller and denser nuclei with a
bigger neutron skin: the Coulomb force pushes
the protons to the surface as already mentioned.
It is also seen that the binding energy is 
always larger for the 
systems considered
without the Coulomb interaction, in the sense that a more 
negative binding 
energy reflects a more bound state. This fact is easily 
explained since the
Coulomb interaction tends to move protons apart and hence, 
diminishes the
binding energy. It is however, interesting to notice that 
the excitation energy increases faster with
temperature when the Coulomb energy is taken into account.
Being a long range force one could think that there would be 
essentially no influence on the excitation energy. In fact, 
for the full calculation, one has $\varepsilon^*(T) = 
\varepsilon(T) - \varepsilon(0) \approx 
\varepsilon_{N}(T) - \varepsilon_{N}(0)$, 
which is
explained by the long range of the Coulomb force.
If the calculation is performed with no Coulomb force ($NC$), the 
excitation energy $\varepsilon^{*\,{NC}}(T) =
\varepsilon^{{NC}}_{N}(T) -
\varepsilon^{NC}_{N}(0)$ contains only the 
nuclear contribution.
In figure \ref{fig3}, the effect of the electromagnetic force 
is clear: if the Coulomb field is taken into account 
the excitation energy is higher, independently 
of the proton-neutron asymmetry, i.e. 
$\varepsilon^{*\,{{NC}}}(T)<\varepsilon^*(T)$. Indeed, we should  
not forget that the Coulomb force affects the proton distribution 
function and, therefore, all properties of the nucleus.
As already mentioned before, we remind that in tables
3 and 5, the last temperature ($6.5$ MeV) is the one at 
which the droplets 
cease to exist within our framework, since we are not able to obtain 
convergence for a droplet of the size considered at higher 
temperatures; the
same is not true in tables 2 and 4, where droplets can 
still be found for temperatures larger than $9$ MeV.

In tables 6 and 7, we compare for the 
nuclei $^{150}_{62}{\rm Sm}_{88}$ and $^{166}_{62}{\rm Sm}_{104}$, 
respectively, the results obtained for the 
quantities already mentioned above, when different parameterizations 
are used. We have considered three values of the temperature and included
the Coulomb interaction.
As previously noticed for the parameterization NL1, 
also for the other two parameterizations, 
the surface energy, the central density and the binding energy per 
particle decrease with the increase of the temperature, while the proton 
radius and the excitation energy per particle increase, under these
conditions.

At $T=0$, the results given by all the parameterizations considered are quite similar. This
reflects the fact that all the parameterizations have been fitted to the  ground-state
properties of stable (NL1) or stable and unstable (NL3 and TM1) 
nuclei. However, for finite
temperatures some differences must be referred.
Although at $T=0$ the surface energy is essentially the same for the 3 parameterizations, it is
seen that for NL1 it reduces much faster with $T$, while in the TM1 it suffers the smallest
reduction. This is true for both proton fractions, 
$Y_p=0.41$ and $0.37$. 
In the last case,
however, $\sigma$ reduces faster with $T$ for all parameterizations.  

The {\it neutron skin thickness}, $\Theta$, is not  sensitive to 
the temperature, but it is interesting to observe that, for a given 
$T$, it assumes smaller values for NL3 and TM1 than for NL1. This 
behaviour was already discussed in \cite{ring} for $T=0$ and is 
 related with the smaller values predicted to the 
symmetry energy for nuclear matter in NL3 (37.4 MeV) and in TM1 (36.9 
MeV) as compared to NL1 (43.5 MeV). Also related with 
these values are the slightly higher binding
energies and the slowlier increasing of the excitation energies with 
temperature for the NL3 and TM1 parameterizations. 

\begin{figure}
\begin{center}
\epsfig{file=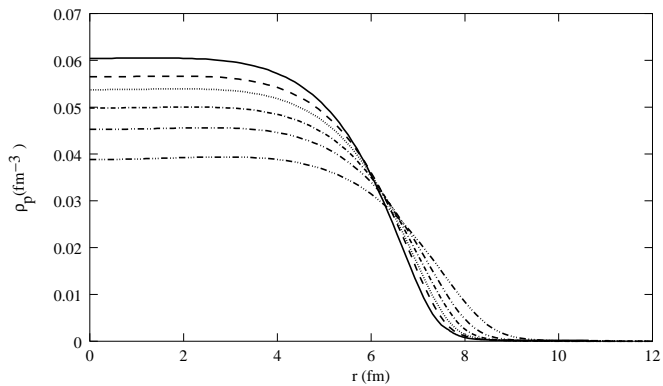,height=10cm,angle=270}
\caption{Density profiles for the protons $\rho_p(r)$  in fm$^{-3}$ are
plotted for $T=5$ MeV and $Y_p=0.44$ (solid line),
$Y_p=0.41$ (dashed line),
$Y_p=0.39$ (dotted line), $Y_p=0.37$ (dot-dashed line), $Y_p=0.35$
(double dot-dashed line) and $Y_p=0.33$ (three dot-dashed line).}
\label{fig4}
\end{center}
\end{figure}

In tables 8 and 9 several properties  of isotopes with $Z=62$ are listed for 
$T=0$ and $T=5$ MeV, respectively. One conclusion 
can immediately be taken: the proton-neutron asymmetry 
makes the nuclei
softer at higher temperatures and properties like the surface energy, the 
central density or the proton
distribution radius vary much more with the number of neutrons if 
temperature is different from zero. It is important that a 
parameterization of the surface energy with temperature  
takes into account the isospin dependence.
The decrease of the central density with the decrease of the proton 
fraction is typical of relativistic mean-field models. 
This feature is different from the prediction of the Skyrme-Hartree-Fock
calculations \cite{LKW97}, where the central 
density does not change with the
proton fraction. We confirm that temperature has no effect on the 
neutron skin.
In figure \ref{fig4}, we show the proton density profiles for 
$Z=62$ and different proton fractions, at $T=5$ MeV. We observe that 
the proton central density increases with the proton fraction. 

\begin{figure}
\begin{center}
\epsfig{file=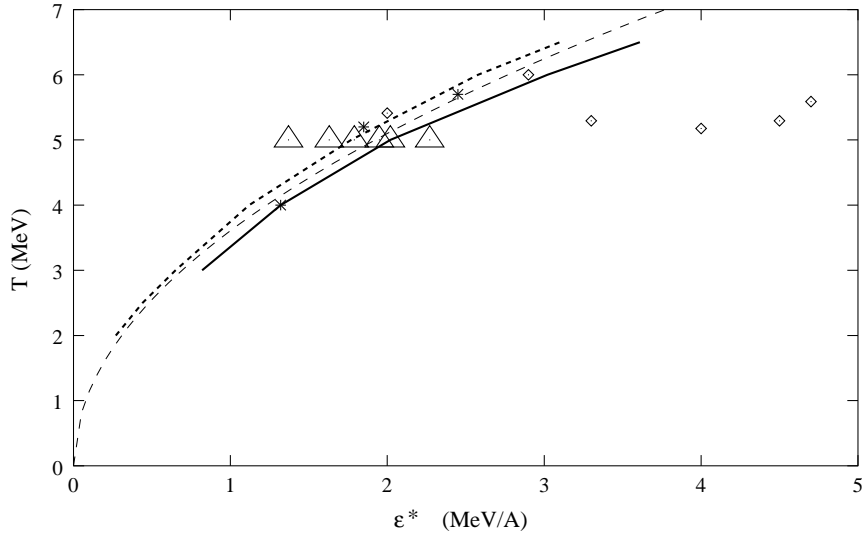,height=13cm,angle=270}
\caption{The caloric curves are shown for  
$^{166}_{62}${\rm Sm}$_{104}$ ($Y_p=0.37$ 
- thick dashed line) and
$^{150}_{62}${\rm Sm}$_{88}$ ($Y_p=0.41$ - thick full line);
the excitation
energies at $T=5$ MeV, for different
proton fractions (big triangles, from
the left to the right $Y_p=0.33, 0.35, 0.37, 0.39, 0.41$ and
$0.44$), the experimental
data from \cite{exp1} (diamonds) and \cite{hagel} (stars), and
the Fermi-gas law \cite{hagel} ($k=13.0$ - thin dashed line) are also 
displayed.}
\label{fig5}
\end{center}
\end{figure}

In figure \ref{fig5} we show the caloric curves obtained 
with the Coulomb interaction, for
$^{150}_{62}{\rm Sm}_{88}$ (thick full line)
and $^{166}_{62}{\rm Sm}_{104}$ (thick dashed line), 
the excitation 
energies at $T=5$ MeV, for different 
proton fractions (big triangles, from 
the left to the right $Y_p=0.33, 0.35, 0.37, 0.39, 0.41$ and 
$0.44$), the experimental 
data from \cite{exp1} (diamonds) and \cite{hagel} (stars), and 
the Fermi-gas law $\varepsilon^*= 
1/k\,T^2$, with
$k=13.0$ (thin dashed line).
We conclude that the excitation energy for $^{166}$Sm (thick dashed 
curve), proton fraction $0.37$, increases slowlier with 
temperature than for $^{150}$Sm (thick full
curve), proton fraction $0.41$. These two curves are consistent with 
data of \cite{hagel} and a level density parameter
$A/k,\, k=13.0$ in the Fermi gas  model relation.
The calculation at $T=5$ MeV, for the proton fractions 
represented in the figure, shows that the caloric
curve is sensitive to the proton-neutron ratio in the compound 
nucleus.

\section {Conclusions}

In summary, we have studied the surface properties and the excitation energies
of arising droplets in a vapor system for temperatures up to 6.5 MeV. The
droplets are described in terms of a Walecka--type model within the
Thomas--Fermi approximation. 
It was shown that, for a fixed temperature, nuclei have a softer 
behaviour with the decrease of its proton fraction, namely the surface 
energy and the central density are smaller and the proton radius becomes 
larger. We have also concluded that the neutron skin is independent of 
the temperature.
Another important conclusion refers to the effect of the Coulomb 
interaction: it reduces the central density of nuclei, as well as the 
neutron skin, since the protons are pushed out to the surface, and 
increases the excitation energy.
We have also compared the behaviour with temperature of
the results obtained within three different 
parameterizations of the non-linear Walecka model, as previously 
referred, NL1 \cite{sed}, 
NL3 \cite{nl3}, TM1 \cite{tm1}. It was shown that, 
at $T=0$, the three parameterizations give similar results but, at 
finite temperature, some of the properties investigated exhibit 
different behaviours: in the NL1 parameterization, the nuclei reacts in 
a softer way to temperature, while in the TM1 parameterization the 
properties of nuclei change slowlier with temperature.
The excitation energies of droplets either corresponding to
$^{150}$Sm or  $^{166}$Sm, for temperatures between 3 and 6.5 MeV, are 
consistent with  the caloric curve in  the Fermi gas approximation with 
a level density parameter  $A/13$. This result 
agrees with experimental 
data obtained in heavy-ion collisions at intermediate energies 
\cite{hagel}. We show that the caloric curve is sensitive 
to the proton fraction  and therefore to the symmetry term of the model 
used. It is also shown that 
close to the critical 
temperature the three parameterizations tested give different results.
Experimentally the dependence on the proton fraction could be studied by 
comparing data obtained from sources with different proton fractions.

\vskip 0.35in
\begin{center}
{\bf Acknowledgments}
\end{center}
This work was partially supported by CNPq - Brazil and CFT - 
Portugal under the contract POCTI/35308/FIS/2000.
The facilities offered by the Center for Computational Physics, University 
of Coimbra are warmly acknowledged.

\newpage
\begin{landscape}

\begin{table}
\centerline{{\bf Table 1.} Sets of parameters used in this work. 
All masses are given in MeV.} 
\vspace{0.5cm}
\begin{center}
\begin{tabular}{|c|c|c|c|c|c|c|c|c|c|c|c|c|c|}
\hline
Force & [Ref.] & $M$  & $m_s$ & $m_v$ & $m_{\rho}$ & $g_s$ & $g_v$ & 
$g_{\rho}$ & $\kappa/M$ & $\lambda$ & $\xi$  \\
\hline
NL1 & \cite{sed} & 938. & 492.25 & 795.36 & 763.0 &
10.138 & 13.285 & 9.952 & 5.122 & -217.613  & 0.0 \\
\hline
NL3 & \cite{nl3} & 939. & 508.194 & 782.501 & 763.0 & 
10.217 & 12.868 & 8.948 &  4.377 & -173.31 & 0.0 \\
\hline
TM1 & \cite {tm1} & 938. & 511.198 & 783.0 & 770.0 & 
10.0289 & 12.6139 & 9.2644 & 3.04 & 3.7098 & 0.0169 \\
\hline
\end{tabular}
\end{center}
\end{table}
\end{landscape}

\begin{table}
\begin{center}
{\bf Table 1a} Nuclear matter properties in the context of several
parameterizations.
\vspace{0.5cm}

\begin{tabular}{|c|c|c|c|c|c|c|c|c|c|c|}
\hline
   &  NL1 \cite{sed}  &  NL3 \cite{nl3}  &  TM1 \cite{tm1}\\
\hline
$B/A$ (MeV) & 16.5 & 16.3 & 16.3 \\
\hline
 
$\rho_0$ (fm$^{-3}$) &  0.153 & 0.148 & 0.145 \\
\hline
 
$K$ (MeV) &  211 & 272 & 281 \\
\hline
 
$a_{sym.}$ (MeV)  & 43.7 & 37.4 & 36.9 \\
\hline
 
$M^*/M$  & 0.57 & 0.60 & 0.63 \\
\hline
\end{tabular}
\end{center}
\end{table}

\newpage
\begin{table}
{\bf Table 2.} 
Output results given by the solution of the coupled differential
equations for different temperatures without the inclusion of the
electromagnetic field for $^{150}_{62}{\rm Sm}_{88}$ ($Y_p=0.41$)
with the NL1 parameterization.
\vspace{0.5cm}

\begin{center}
\begin{tabular}{|c|c|c|c|c|c|c|c|c|c|c|c|}
\hline
$T$&  $\sigma$ & $R_p$ & $\Theta$ &$\rho(0)$& $B/A$ &  $\varepsilon^*(T)$
\\
(MeV) & (MeV fm$^{-2}$) & (fm)&  (fm)&(fm$^{-3}$)& (MeV/A) & (MeV/A) 
\\ 
\hline
\hline
0.   & 1.15 & 5.88 & 0.33 &0.161& -11.82  & 0.00  \\
3.   & 1.14 & 5.77 & 0.39 & 0.166&-10.95  & 0.87 \\ 
4.   & 1.06 & 5.84 & 0.37 & 0.163&-10.64  & 1.18 \\
5.   & 0.95 & 5.92 & 0.36 &0.156& -10.14  & 1.68 \\
6.   & 0.83 & 6.04 & 0.35 & 0.147& -9.49 & 2.33 \\
6.5 & 0.74 & 6.13 & 0.35 &0.140&  -8.99 & 2.83 \\
7.   & 0.68 & 6.22 & 0.35 & 0.137& -8.59  & 3.23 \\
8.   & 0.54 & 6.41 & 0.34 &0.126&  -7.56  & 4.26 \\ 
9.   & 0.40 & 6.63 & 0.36 &0.112 & -6.33 & 5.49 \\
\hline 
\end{tabular}
\end{center}
\end{table}

\begin{table}
{\bf Table 3.} 
Output results given by the solution of the coupled differential
equations for different temperatures with the inclusion of the
electromagnetic field for $^{150}_{62}{\rm Sm}_{88}$ ($Y_p=0.41$)
with the NL1 parameterization. 

\begin{center}
\begin{tabular}{|c|c|c|c|c|c|c|c|c|c|c|c|c|c|}
\hline
$T$&  $\sigma$ & $R_p$ & $\Theta$ &$\rho(0)$& $B/A$ 
& $\varepsilon^*(T)$
\\
(MeV) & (MeV fm$^{-2}$) & (fm)&  (fm)& (fm$^{-3}$&(MeV/A) & (MeV/A) 
\\ 
\hline
\hline
0.   & 1.00 & 6.22 & 0.16 & 0.142& -8.39  & 0.00 \\
3.   & 0.98 & 6.23 & 0.18 & 0.140& -7.56  & 0.83  \\
4.   & 0.91 & 6.30 & 0.17 & 0.137& -7.06  & 1.33  \\
5.   & 0.81 & 6.40 & 0.16 & 0.131& -6.37  & 2.02 \\
6.   & 0.65 & 6.57 & 0.15 & 0.121& -5.38  & 3.01 \\
6.5  & 0.57 & 6.68 & 0.14 & 0.115& -4.78  & 3.61 \\
\hline 
\end{tabular}
\end{center}
\end{table}

\begin{table}
{\bf Table 4.} 
Output results given by the solution of the coupled differential
equations for different temperatures without the inclusion of the
electromagnetic field for $^{166}_{62}{\rm Sm}_{104}$ ($Y_p=0.37$)
with the NL1 parameterization.

\begin{center}
\begin{tabular}{|c|c|c|c|c|c|c|c|c|c|c|c|}
\hline
$T$&  $\sigma$ & $R_p$ & $\Theta$ &$\rho(0)$& $B/A$ 
& $\varepsilon^*(T)$
\\
(MeV) & (MeV fm$^{-2}$) & (fm)&  (fm)&(fm$^{-3}$)& (MeV/A) & (MeV/A)
\\ 
\hline
\hline
0.   & 1.03 & 6.02 & 0.51 & 0.157&-11.03 & 0.00 \\ 
2.   & 0.99 & 6.03 & 0.53 & 0.154&-10.71  & 0.32  \\
3.   & 0.92 & 6.08 & 0.54 & 0.149&-10.43 & 0.60  \\
4.   & 0.84 & 6.16 & 0.54 & 0.144&-10.12  & 0.91  \\
5.   & 0.74 & 6.25 & 0.52 & 0.138& -9.70  & 1.33  \\
6.   & 0.61 & 6.39 & 0.53 & 0.129& -9.12  & 1.91 \\
7.   & 0.49 & 6.57 & 0.53 & 0.120& -8.44 & 2.59 \\
8.   & 0.35 & 6.87 & 0.55 & 0.105& -7.57  & 3.46 \\
9.   & 0.27 & 7.08 & 0.52 & 0.098& -6.46  & 4.57 \\
\hline 
\end{tabular}
\end{center} 
\end{table}

\begin{table}
{\bf Table 5.} 
Output results given by the solution of the coupled differential
equations for different temperatures with the inclusion of the
electromagnetic field for $^{166}_{62}{\rm Sm}_{104}$ ($Y_p=0.37$)
with the NL1 parameterization.

\begin{center}
\begin{tabular}{|c|c|c|c|c|c|c|c|c|c|c|c|c|}
\hline
$T$&  $\sigma$ & $R_p$ & $\Theta$ &$\rho(0)$& $B/A$ 
& $\varepsilon^*(T)$
\\
(MeV) & (MeV fm$^{-2}$) & (fm)&  (fm)&(fm$^{-3}$)& (MeV/A) & (MeV/A) 
\\ 
\hline
\hline
0.    & 0.92 & 6.40 & 0.32 &0.138  & -7.93  & 0.00 \\
2.   & 0.91 & 6.42 & 0.31 & 0.137  & -7.65  & 0.28 \\
3.   & 0.86 & 6.46 & 0.32 & 0.134  & -7.26  & 0.67 \\
4.   & 0.77 & 6.54 & 0.33 & 0.129  & -6.80  & 1.13 \\
5.   & 0.66 & 6.67 & 0.32 & 0.122  & -6.14  & 1.79 \\
6.   & 0.53 & 6.87 & 0.31 & 0.113  & -5.20  & 2.73 \\
6.5  & 0.47 & 7.00 & 0.29 & 0.108  & -4.81  & 3.12\\
\hline 
\end{tabular}
\end{center} 
\end{table}

\begin{landscape}
\begin{table}
{\bf Table 6.} Output results given by the solution of the coupled
differential equations with the inclusion of the electromagnetic
field, for $^{150}_{62}{\rm Sm}_{88}$ ($Y_p$=0.41). For different
temperatures, the results obtained with several parameterizations
are shown.
\vspace*{0.5cm}
\begin{center}
\begin{tabular}{lccccccccccc}
\hline
\hline
&  & T=0 MeV&  &$\phantom{33}$ & &  T=5 MeV & & $\phantom{33}$  & & T=6
MeV&\\
\cline{2-4} \cline{6-8} \cline{10-12}
    & NL1 & NL3 & TM1  && NL1 & NL3 & TM1 && NL1 & NL3 & TM1 \\
 
\hline
$\sigma$ (MeV fm$^{-2}$) & 1.000 & 1.007 & 0.987 &&
0.810 & 0.857 & 0.874 && 0.650 & 0.725 & 0.762 \\
\hline
 
$R_p$ (fm) & 6.22 & 6.26 & 6.30 && 6.40 & 6.37 & 6.39 &&
6.57 & 6.49 & 6.50 \\
\hline
 
$\Theta$ (fm) & 0.16 & 0.11 & 0.09 && 0.16 & 0.12 & 0.10 &&
0.15 & 0.11 & 0.10\\
\hline
 
$\rho(0)$ (fm$^{-3}$) & 0.142 & 0.142 & 0.139 &&
0.131 & 0.134 & 0.133 && 0.121 & 0.127  & 0.127 \\
\hline
$B/A$ (MeV/A) & -8.39 & -8.43 & -8.52 && -6.37 & -6.49 & -6.66 &&
 -5.38  & -5.64 & -5.85\\
\hline
 
$\varepsilon^{\star}(T)$ (MeV/A) & 0.00 & 0.00 & 0.00 &&
2.02 & 1.93 & 1.87 && 3.01 & 2.79 & 2.67\\
\hline
\end{tabular}
\end{center} 
\end{table}
\end{landscape}

\begin{landscape}
\begin{table}
{\bf Table 7.} Output results given by the solution of the coupled
differential equations with the inclusion of the electromagnetic
field, for $^{166}_{62}{\rm Sm}_{104}$ ($Y_p$=0.37). For different
temperatures, the results obtained with several parameterizations
are shown.
\vspace*{0.5cm}
\begin{center}
\begin{tabular}{lccccccccccc}
\hline
\hline
&  & T=0 MeV&  &$\phantom{33}$ & &  T=5 MeV & & $\phantom{33}$  & & T=6
MeV&\\
\cline{2-4} \cline{6-8} \cline{10-12}
    & NL1 & NL3 & TM1  && NL1 & NL3 & TM1 && NL1 & NL3 & TM1 \\
\hline
$\sigma$ (MeV fm$^{-2}$) & 0.920 & 0.933 & 0.917 &&
0.660 & 0.723 & 0.742 && 0.530 & 0.615 & 0.640 \\
\hline
 
$R_p$ (fm) & 6.40 & 6.43 & 6.49 && 6.67 & 6.62 & 6.66 &&
6.87 & 6.75 & 6.76 \\
\hline
 
$\Theta$ (fm) & 0.32 & 0.25 & 0.22 && 0.32 & 0.26 & 0.23 &&
0.31 & 0.25 & 0.22\\
\hline
 
$\rho(0)$ (fm$^{-3}$) & 0.138 & 0.139 & 0.137 &&
0.122 & 0.128 & 0.127 && 0.113 & 0.120 & 0.122\\
\hline

$B/A$ (MeV/A) & -7.93& -8.04 & -8.11  && -6.14 & -6.30 & -6.43 &&
-5.20 & -5.50 & -5.67\\
\hline
 
$\varepsilon^{\star}(T)$ (MeV/A) & 0.00 & 0.00 & 0.00 &&
1.79 & 1.74 & 1.68 && 2.73 & 2.54 & 2.45\\
\hline
\end{tabular}
\end{center}
\end{table}
\end{landscape}
   
\newpage
\begin{table}
{\bf Table 8.} 
Output results given by the solution of the coupled differential
equations for $T=0$ with the inclusion of the
electromagnetic field for $Z={62}$ and different number of neutrons
with the NL1 parameterization.
\begin{center}
\begin{tabular}{|c|c|c|c|c|c|c|c|c|c|c|c|c|}

\hline
$N$&  $Y$  &$\sigma$& $R_p$ & $\Theta$ &$\rho(0)$& $B/A$ 
\\
& & (MeV fm$^{-2}$) & (fm)&  (fm)&(fm$^{-3}$)& (MeV/A) 
\\ 
\hline
\hline
78 & 0.445 & 1.05 & 6.11& 0.05&   0.144 & -8.59  \\
88 & 0.413 & 1.00 & 6.22 & 0.16 & 0.142&-8.39 \\
94& 0.397  & 0.97 & 6.29& 0.22&  0.141 &-8.25\\
104&0.374  & 0.92 & 6.40 & 0.32 &0.138  & -7.93 \\
112 &0.358 & 0.87 & 6.50& 0.39& 0.136&-7.76\\
124& 0.334 & 0.80 & 6.62 & 0.50&  0.133  &-7.37\\
\hline 
\end{tabular}
\end{center} 
\end{table}

\begin{table}
{\bf Table 9.} 
Output results given by the solution of the coupled differential
equations for $T=5$ with the inclusion of the
electromagnetic field for $Z={62}$ and different number of neutrons
with the NL1 parameterization.

\begin{center}
\begin{tabular}{|c|c|c|c|c|c|c|c|c|c|c|c|c|c|}
\hline
$N$&  $Y$  &$\sigma$& $R_p$ & $\Theta$ &$\rho(0)$& $B/A$&
$\varepsilon^*(T)$
\\
& & (MeV fm$^{-2}$) & (fm)&  (fm)&(fm$^{-3}$)& (MeV/A) & (MeV/A) 
\\ 
\hline
\hline
78& 0.445 & 0.87 & 6.26& 0.04 & 0.134 & -6.32& 2.27 \\
88 & 0.413 & 0.81 & 6.40& 0.16 & 0.131 & -6.37&2.02  \\
94 & 0.396 & 0.75 & 6.46 & 0.23 & 0.128 & -6.32& 1.93 \\
104& 0.374 & 0.66 & 6.67 & 0.32 & 0.122 & -6.14 & 1.79\\
112 & 0.355 & 0.55 & 6.89 & 0.39 & 0.115 & -6.13& 1.63 \\
124 & 0.334 & 0.41 & 7.27 & 0.48 & 0.102&-6.01 & 1.37 \\
\hline 
\end{tabular}
\end{center} 
\end{table}

\end{document}